\documentclass{article}

     \PassOptionsToPackage{numbers, compress}{natbib}
     \usepackage[preprint]{neurips_2020}

\usepackage[utf8]{inputenc} % allow utf-8 input
\usepackage[T1]{fontenc}    % use 8-bit T1 fonts
\usepackage[hidelinks]{hyperref}       % hyperlinks
\usepackage{url}            % simple URL typesetting
\usepackage{booktabs}       % professional-quality tables
\usepackage{amsfonts}       % blackboard math symbols
\usepackage{nicefrac}       % compact symbols for 1/2, etc.
\usepackage{microtype}      % microtypography
\usepackage{natbib}
\usepackage{graphicx}
\usepackage{textcomp}
\usepackage{xcolor}
\usepackage{algorithm}
\usepackage{algpseudocode}
\usepackage{amsmath, bm}
\usepackage{hyperref}
\usepackage{setspace}

\title{Duplicate Question Retrieval and Confirmation Time Prediction in Software Communities}

\author{
Rima Hazra, Debanjan Saha, Amruit Sahoo, Somnath Banerjee, Animesh Mukherjee\\
Indian Institute of Technology Kharagpur\\ 
\texttt{ \{to\_rima, debanjansaha, amruit2k\}@iitkgp.ac.in}\\
\texttt{ som.iitkgpcse@kgpian.iitkgp.ac.in, animeshm@cse.iitkgp.ac.in} 
}

\begin{document}

\maketitle

\begin{abstract}
Community Question Answering (CQA) in different domains is growing at a large scale because of the availability of several platforms and huge shareable information among users. With the rapid growth of such online platforms, a massive amount of archived data makes it difficult for moderators to retrieve possible duplicates for a new question and identify and confirm existing question pairs as duplicates at the right time. This problem is even more critical in CQAs corresponding to large software systems like askubuntu where moderators need to be experts to comprehend something as a duplicate. Note that the prime challenge in such CQA platforms is that the moderators are themselves experts and are therefore usually extremely busy with their time being extraordinarily expensive. To facilitate the task of the moderators, in this work, we have tackled two significant issues for the askubuntu CQA platform: (1) retrieval of duplicate questions given a new question and (2) duplicate question confirmation time prediction. In the first task, we focus on retrieving duplicate questions from a question pool for a particular newly posted question. In the second task, we solve a regression problem to rank a pair of questions that could potentially take a long time to get confirmed as duplicates. For duplicate question retrieval, we propose a Siamese neural network based approach by exploiting both text and network-based features, which outperforms several state-of-the-art baseline techniques. Our method outperforms DupPredictor~\cite{Zhang:2015} and DUPE~\cite{Ahasanuzzaman:2016} by 5\% and 7\% respectively. For duplicate confirmation time prediction, we have used both the standard machine learning models and neural network along with the text and graph-based features. We obtain Spearman's rank correlation of 0.20 and 0.213 (statistically significant) for text and graph based features respectively.
\end{abstract}

\maketitle

\section{Introduction}
Community question answering (CQA) platforms are rapidly becoming popular because of their extensive collection of questions and answers. Due to the burgeoning growth of such CQA portals, questions posted by users can be repetitive. In many cases, new users tend to post duplicate questions since they are not fully aware of the navigation tools available on the platform. Moderators/experienced users need to identify and mark duplicate questions in such cases. This becomes extremely challenging and time-consuming given the scale of data they need to sieve through. While posting a new question, if a user is prompted with similar (or precisely the same) queries reported previously, it can reduce the platform's redundancy. CQAs pertaining to large software systems like askubuntu pose a larger challenge since the moderators need to be mostly experts to identify if a question is a duplicate. The availability of such experts is limited and usually quite expensive.
%identifying
Further confirming a pair of questions as actual duplicate is a manual (mostly moderator or experienced users) task. The manual nature of this task leads to the consumption of a long time (with respect to the speed of knowledge exchange in the community) for a pair of questions to get confirmed as duplicate since it was the first identified. For instance, as per the askubuntu policy, at least five votes are needed to confirm that a pair of questions are duplicates. Typically, these votes get accrued over a long period of time and increase the time to closure. 
%solve duplicate question retrieval
In this work, we attempt to retrieve possible duplicate questions for a newly posted question. Further, we attempt to direct the moderator's attention toward marked duplicate pairs that could have got identified (confirmed) in longer than usual time. We will use queries and questions interchangeably in the following sections.\\
\noindent \textbf{Duplicate question retrieval}: In this task, for each new query, we shall attempt to recommend the top $k$ possible duplicates to the users so that they have the option to choose a similar query from the previously posted questions. 
%This experiment will have the advantage of identifying the exact/similar problem from the previous popular posts by only writing some core words. 
When a new user posts a repetitive question, a moderator should be able to quickly find the possible duplicates from the earlier queries. An example of redundant question is noted in Table~\ref{tab:dupQuesEx}.\\
\noindent \textbf{Duplicate confirmation time prediction}: It is observed that after a pair of queries are initially marked as a possible duplicate, it takes a long time for them to acquire enough votes to be eventually confirmed as duplicates. In askubuntu, as per our analysis, there are around $\sim 40\%$ question pairs that take more than five days to get confirmed as duplicates. Also, out of all these pairs, 50-55\% have a high view count of 1000 -- 10,000 thus showing that they engage a lot of users. Our task is to identify those pairs which took a long time to be confirmed as duplicates. We intend to get a rank list of the pairs according to their {\em time taken} in decreasing order. Such pairs will be explicitly suggested to the moderators for more attention.
%~\rh{There are a few reasons for posing this task. In every platform, there are certain rules present to identify and confirm duplicate pairs. In Askubuntu, If the duplicate pairs are marked as duplicate by k number of users (with high reputation score), then only the pair get confirmed as duplicate. Now, there are around ~40\% pairs in askubuntu data which takes more than 5 days to get confirmed as duplicate. Also, out of all those pairs, 50-55\% question pairs has high view count more than 1000 and 10,000). 
%But for new portals, not always possible to get experienced users who can mark a pair as duplicate. Here, we are reducing dependency on the experienced users of the system for confirming a pair as duplicate. So, we want to make an automated system which will help the moderator to identify the pairs of marked duplicates with longer time to get confirmed. Provide average time required to obtain five votes.} 
An example of duplicate question pairs and their duplicate confirmation timestamp is noted in Table~\ref{tab:dupQuesEx}.
\begin{table}[h]%[!htb]
%\vspace*{-.7cm}
\tiny
\centering
%\begin{minipage}{\textwidth}
%p{4.5cm}|p{4.5cm}|p{2.3cm}
\resizebox{.75\textwidth}{!}{
\begin{tabular}{|p{3cm}|p{3cm}|p{1cm}|} \hline 
\centering {\bf Question A} &  \centering {\bf Question B} & {\bf DCT} \\ \hline
 \textbf{Title}: How to remove WUBI installed Ubuntu without affecting Windows files? & \textbf{Title}: How do I uninstall Ubuntu Wubi? & \\
 \cline{1-2}
 \textbf{Body}: \dots Now I want to remove Ubuntu from my laptop without affecting my Windows files \dots & \textbf{Body}: I want to uninstall Ubuntu because I just don't like it...  I have windows 7 \dots & 2013-02-18 03:03:21 \\ 
\cline{1-2}
\textbf{Posting time}: 2012-11-07 13:35:45 & \textbf{Posting time}: 2012-05-30 18:58:11 & \\ \hline
\end{tabular}
}
\caption{ Duplicate confirmation timestamp. Example of a pair of duplicate questions with the title, body and posting timestamp. The duplicate link formation is also given.}%\footnotesize
%\end{minipage}
\label{tab:dupQuesEx}
%\vspace*{-1.18cm}
\end{table}
To the best of our knowledge, the first problem, i.e., duplicate question retrieval, has been treated as a classification task~\cite{Wang:2020,Pei:2021, bogdanova-etal-2015-detecting} or as a recommendation task using a classification/regression objective function~\cite{Ahasanuzzaman:2016, Zhang:2018}.
%cross-encoder model
However, such a scheme cannot be easily deployed in a real-time scenario given a large corpus. In this work, we treat this problem as a recommendation task and propose a method based on Siamese neural network~\cite{Bromley:1993} to solve the problem. In addition, we use the node embedding obtained from the tag co-occurrence network as the representation of a tag in order to enrich our model. Further, we compare the state-of-the-art methods~\cite{Ahasanuzzaman:2016,Zhang:2015, Wang:2020, Robertson:2009} with our approach. The second problem, i.e., duplicate confirmation time prediction, has not been attempted in literature for any platform to the best of our knowledge. However, this problem is important when the system already has a lot of unconfirmed duplicate pairs.\\
% To the best of our knowledge, the first problem, i.e., duplicate question retrieval, is that most earlier works have treated this as a classification task~\cite{Pei:2021}~\cite{Wang:2020} and recommendation using cross encoder model which cannot be easily deployed in a real-time scenario with large corpora. In this work, we treated this problem as a recommendation task and proposed a Siamese network model~\rh{remove wiki link and add research paper}~\footnote{} to solve the problem. The second problem, i.e., duplicate identification time prediction, has not been attempted in literature for any platform. \\
%\todo[inline,color=yellow!15]{Briefly tell about our work and notable results}
\noindent \textbf{Our contribution and results}:\\
\noindent \textit{Duplicate question retrieval}: We propose a simple method for retrieving actual duplicates from the candidates of a given question. We compare our approach with various state-of-the-art baselines. First, we use question title and body text representation as features. Further, including tag representations from the tag co-occurrence network increases the overall performance. Using text features, we obtain an MRR of 9.45\%, considering a list of 485 duplicates with an average candidate set size of 5941. In addition, the recall rate RR@$10$ is 15.88\%. The inclusion of network features brings additional benefits, which leads the MRR and RR@$10$ to rise to 11.10\% and 18.35\% (considering a list of 485 duplicates with an average candidate set of size 5941), respectively. Our model's uniqueness lies in tackling this problem by \textit{not} using a conventional classification objective function~\cite{Ahasanuzzaman:2016,Zhang:2018}
%conventional cross-encoder~\cite{}\am{cite?} 
and how we sample the negative examples and select the candidate set. Further, including features from the tag co-occurrence network along with textual features helps to considerably outperform the baseline approaches. \\
\noindent \textit{Duplicate confirmation time prediction}: We model this problem as a regression task where the input is the text representation of a question (aka text), and the output is a probability ranking of questions (based on the time required to close a question as duplicates from the time they have been first identified as being duplicates). Including tag representations obtained from the tag co-occurrence network as additional features (aka text+network) further improves the performance. MLP-based models achieve the best Spearman's rank correlation of 0.208 (text) and 0.213 (text+network), respectively, considering the complete rank list of 3756 duplicates. Adding network features always shows improvement, and these results are statistically significant. 
While we perform our experiments on the askubuntu platform, we would like to highlight that our methods are generic and can be extended to any other platform.

\section{Related work}
\if{0}\noindent \textbf{Characterization of CQA systems}: Various studies have been conducted to characterize different aspects of the CQA platforms. Such characterization involves answerability prediction~\cite{Dror:2013,Maity:2018,Yazdaninia:2021}, dynamics of interaction~\cite{Mamykina:2011,Qu:2018}, topic modelling and question retrieval~\cite{Chen:2016}, recommending experts to the question~\cite{Yang:2014,Wang:2016,Mumtaz:2019,Nikzad:2021}, 
characteristics of closed questions~\cite{Correa:2013}, duplicate question detection~\cite{Ahasanuzzaman:2016}, time taken to close the duplicate questions~\cite{Ahasanuzzaman:2016}, characterizing and modelling the deleted questions~\cite{Correa:2014}, analyzing interactions in duplicate questions~\cite{Rodrigues:2017}, answer selection and recommendation~\cite{Quan:2015,Xiaoqiang:2018,Nakov:2019}, question quality assessment~\cite{Baltadzhieva:2015} and automatic moderation action on the CQA website~\cite{Annamoradnejad:2020}. \\\fi
\noindent \textbf{Duplicate question retrieval}: 
Duplicate detection is one of the major problems in various large systems since the growth of Internet usage. Duplicate detection has been an important problem in databases~\cite{Yang:2006,Gong:2008}, webs~\cite{Yandrapally:2020}, bug tracking systems~\cite{Runeson2007DetectionOD,Sun:2011,Alipour:2013, 10.1007/978-3-031-26422-1_15}, and community question answering systems~\cite{Zhang:2015,Ahasanuzzaman:2016,Zhang:2017WWW,Prabowo2019DuplicateQD}. %There is also a lot of existing research~\cite{Roul:2020,Luciv2018}  for detecting near-duplicate documents.
%the scale of the dataset is small.
Zhang {\em et al}~\cite{Zhang:2015} proposed a novel method called DupPredictor to identify possible duplicates of a new question by considering various factors. %They used the title, body and tags of the questions to compute the features. They have computed cosine similarity between titles, bodies, topics and tags for a pair of questions. To obtain the topic, they used LDA topic modelling. Further, they combined these four similarity scores to obtain a new similarity score.
%The authors in~\cite{Ahasanuzzaman:2016} conducted a manual study on the duplicate questions in StackOverflow. They investigated various factors (duplicate questions posted over time, time taken to close the duplicate questions, tag analysis of duplicate questions) for submitting duplicate questions by users. Further, they identified a set of important features and proposed a classification technique for detecting duplicates. We have reimplemented their methodology as an additional baseline and experimented with our dataset.
The authors in~\cite{Zhang:2017WWW} proposed a classification method for duplicate question detection on StackOverflow\footnote{The results could not be reproduced due to lack of requisite information about experimental setup and feature calculation.}. For a pair of questions, they obtain the features from word2vec, topic modelling and phrase pairs that co-occur in duplicate questions. %They have conducted the experiment on StackOverflow data. We could not reproduce their implementation owing to the lack of requisite information about experimental setup and feature calculation.
%\\
% ~\rh{Zhang {\em et al.}~\cite{Zhang:2018}}
%Textual scemantic similarity
In~\cite{bogdanova-etal-2015-detecting} the authors used standard machine learning models such as support vector machine and convolutional neural network to identify semantically similar questions in an online forum. %They observed that CNN with in-domain word embeddings performed best on limited training data.
The authors in~\cite{Kumari:2021} used Siamese-LSTM~\cite{Mueller_Thyagarajan_2016} along with dense layer and classifier to detect semantically equivalent question pairs in Quora. %They have conducted experiments on the Quora dataset.
In~\cite{Homma:2016} the authors used Siamese GRU network to detect the semantically equivalent question pairs in Quora. %They also conducted experiments with different distance measures. 
Finally, the authors in~\cite{Jabbar:2021} proposed a Siamese network based method for detecting duplicate questions in StackExchange data. Further, they employed domain adaptation with transfer learning to improve performance\footnote{In~\cite{Zainab:2020}, although the authors used Siamese neural networks, the duplicate pairs for testing are predefined and thus cannot be used as an additional baseline.}.

% In the last few years, finding duplicate questions and identifying them is becoming challenging due to the scale of growth in the CQA platforms. Authors ~\rh{of earlier research} have attempted to solve this problem on different CQA platforms such as Quora, Yahoo! Answers, AskUbuntu, StackOverflow, etc.  
% Mainly, two types of architecture are used to solve this problem in the literature. In the first type, Siamese network based architecture is used to develop various neural models. In Siamese network architecture, authors encode textual information and use neural networks to generate a fixed-length representation for each question. Further, they use different similarity measures to decide the relatedness between two queries. These methods can be generalized over various tasks. The second type are the methods developed from the cross encoder architecture. In cross encoder-like architecture, cross attention between two queries has been performed to decide the similarity between two questions. In the literature, most of the methods~\cite{Zhang:2018,Shah:2018,Wang:2020} are based on the cross encoder architecture.
% The bi-encoder methods are more suitable and easy to adapt for large datasets. ~\rh{None or very few -- Fix this} bi-encoder-based techniques are present in the literature. Most works employ cross features even at the end of their model architecture. These cross-encoder models are computationally heavy and not suitable for large datasets.\\
\noindent \textbf{Identifying duplicate question time}: Confirming a pair of a question as a duplicate within the tangible time frame is a challenging task. Less or no earlier work is present where this problem has been addressed. In~\cite{Ahasanuzzaman:2016}, while characterizing the same questions in the StackOverflow platform, the authors have analyzed the time taken to close a question as duplicate.\\ %They observed that around 69\% questions took more than one day to be {\em closed} as duplicate, and around 29\% questions were marked {\em closed} within ten hours. Further, they proposed a model to recommend the duplicate questions already answered. \\
Our work is unique in different ways. First, we have considered the latest dump of a popular CQA platform -- askubuntu -- vital for the software development community. While our model is simple, the main novelty lies in how we perform negative sampling and candidate set selection for duplicate retrieval. Further, we conceive of a novel tag co-occurrence network that brings additional performance boosts for both tasks.
\section{Dataset}
%\vspace{-0.2cm}
In this paper, we use the community question-answering platform askubuntu data dump released at the beginning of 2021. The data dump consists of $\sim$366K questions and textual information such as question title, question body, and corresponding answers. Question metadata includes question reporting time, question tags, answer posting time, question reporting user, users who posted the answers, and duplicate link formation timestamp. The primary contents of the dataset are noted in Table~\ref{tab:datastat}. For our experiment, we have chosen askubuntu because it is based on a single ecosystem (ubuntu ecosystem) and contains large volumes of duplicates. Further, moderators on these platforms are experts who are usually very busy with their time being extraordinarily expensive. In previous papers, certain question groups (Java, C++, Python, Ruby, HTML, and objective-C)~\cite{Ahasanuzzaman:2016} or older repo (contains 1641 duplicates only)~\cite{Zhang:2015} of StackOverflow and StackExchange has been used for duplicate detection. The authors of paper~\cite{bogdanova:2015} used the askubuntu data for detecting schematically equivalent questions.
%\fi
% we have chosen askubuntu because --> single topic with large volume. Whereas in stackoverlow, huge amount of diverse data present. and people produced results after seggregating them based on the topic. Which downscale the number of questions belongs to each topic in stackoverflow.}
\begin{table}[!htb]
%\vspace*{-.5cm}
\centering
\tiny
\resizebox{.75\textwidth}{!}{
\begin{tabular}{|p{4cm}|p{2.2cm}|} \hline
{\bf Information} & {\bf Count}  \\ \hline \hline
Total number of questions & 366,324 \\ \hline
Average number of words in body & 139.62\\ \hline
Average number of words in title & 8.51 \\ \hline
Average number of tags &  2.77\\ \hline
Total number of tags &  3118\\ \hline
\#duplicate question pairs & 68286 \\ \hline
\end{tabular}
}
\caption{\footnotesize Dataset statistics.}\label{tab:datastat}
%\vspace*{-1.01cm}
\end{table}

\begin{figure}
%\vspace*{-.5cm}
\centering
\small
\begin{minipage}{.6\textwidth}
\includegraphics[width=1\textwidth]{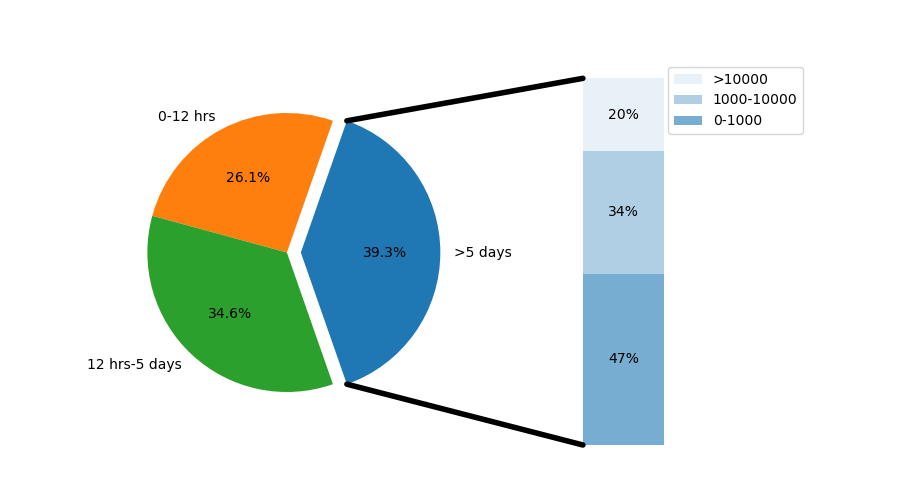} \\
%\vspace{-0.5 cm}
\caption{\footnotesize The pie chart represents the fraction of marked duplicate pairs that have been confirmed as duplicates after a particular time. The corresponding bar shows the view count of the latest question in the pair which took more than 5 days to get confirmed.}  
\label{fig:duptimeFrac}
\end{minipage}
% \begin{minipage}{.23\textwidth}
% \includegraphics[width=0.9\textwidth]{Figures/dup_5GT_viewcount.png} \\
% %\vspace{-0.5 cm}
% \caption{\footnotesize Fraction of duplicate pairs}  
% \label{fig:dupViewCount}
% \end{minipage}
%\caption{\footnotesize Fraction of duplicate pairs} 
\end{figure}

In this paper, we use the community question-answering platform askubuntu data dump released at the beginning of 2021. The data dump consists of $\sim$366K questions and textual information such as question title, question body, and corresponding answers. Question metadata includes question reporting time, question tags, answer posting time, question reporting user, users who posted the answers, and duplicate link formation timestamp. The primary contents of the dataset are noted in Table~\ref{tab:datastat}. Note that we have the tags associated with the questions. Like any other platform, these tags attempt to topically organize the questions to facilitate a better search. %Thus tags have a critical role in the overall CQA ecosystem. 
An inspection of these tags across the duplicate questions show that they are primarily on system configuration requirements for Ubuntu installation, driver installation, new package installation, and suitable Ubuntu distribution according to the hardware configuration and basic Linux commands.\\
In order to extract meaningful information from these tags and their relationships we construct a tag co-occurrence network where the tags are the nodes and two nodes are connected if they co-occur in a question. We compute the Jaccard overlap of question sets to which two tags ${t_1, t_2}$ are common which defines the weight of the edge. Edges with a weight larger than 0.005 are only retained\footnote{We set this threshold based on manual inspection of the data.}.\\ %In this tag co-occurrence network (see~\ref{fig:example_graph} and ~\ref{fig:projected_graph}), we have 3118 nodes with a total number of 20227 edges.
Another essential parameter in our data is the duplicate confirmation time. We assume the pairs are marked as soon as the recent most question of the pair has been posted. We observe that there are $\sim 40\%$ question pairs that require more than 5 days to become confirmed as duplicates (see Fig.~\ref{fig:duptimeFrac}). Around 20\% of these pairs are viewed by more than 10000 users and 30-35\% of these pairs are viewed by 1000-10000 users showing high levels of user engagement thus necessitating the prediction of duplicate question pair confirmation time.\\
%\noindent \textbf{Confirmation of duplicate pairs}:  We categorize all such pairs into three buckets -- (a) the fraction of pairs that are confirmed within 0-12 hours, (b) the fraction of pairs that are confirmed between 12 hours and 5 days, and (c) the fraction of pairs that are confirmed after 5 days. We observe that there are $\sim 40\%$ question pairs that require more than 5 days to become confirmed as duplicates (see Figure~\ref{fig:duptimeFrac}). Next, we compute the level of user engagement in terms of the number of views\footnote{We have considered the view count of the recent most question from each duplicate pairs.} of these $\sim 40\%$ question pairs. Figure~\ref{fig:dupViewCount}, shows the distribution of the view counts of these $\sim 40\%$ question pairs. We observe that 30-35\% of these question pairs have a view count between 1000-10000 while another 20\% of these pairs have a view count larger than 10000. The above study indicates that quite a large number of question pairs with high view counts remain unconfirmed for a large period of time. This underpins the motivation for our second problem on 
For our experiment, we have chosen askubuntu because it is based on a single ecosystem (ubuntu ecosystem) and contains large volumes of duplicates. Further, moderators on these platforms are experts who are usually very busy with their time being extraordinarily expensive.
% \begin{figure}
% %\vspace*{-.5cm}
% %\centering
% \small
% % \begin{minipage}{.47\textwidth}
% \includegraphics[width=0.4\textwidth]{Figures/dup_5GT_viewcount.png} \\
% %\vspace{-0.5 cm}
% \caption{\footnotesize Fraction of duplicate pairs}  
% \label{fig:dupViewCount}
% % \end{minipage}
% \end{figure}

\section{Notation and preliminaries}
We have a set of $Q$ questions in a CQA ecosystem indexed as $q \in [Q] = [1 \dots Q]$ where $q$ represents a single question. Each question $q$ is associated with the reporting timestamp $ts(q)$. There is a set of tags $\mathcal{T}$ indexed by $t \in [\mathcal{T}] = [1 \dots \mathcal{T}]$. Given a question $q$, there is a set of associated tags $T_q \subset \mathcal{T}$. Each question $q$ has three important features -- the title of the question denoted by $\mathcal{QT}$, the body of the question denoted by $\mathcal{QB}$, and the tags of the question denoted by $T_q$. We have a set of duplicate pairs of questions $(q_1, q_2)$ where $q_1, q_2 \in Q$. We assume that the latest question within the duplicate pair is an anchor question $q_1$ and the older question as its master (usually positive pair) denoted by $q_2$. In specific, for a duplicate pair $(q_1, q_2)$, $q_1$ is an anchor if $ts(q_1) > ts(q_2)$; else $q_2$ is the anchor. We denote the time when the question was confirmed by $ts(q_1, q_2)$ where ($q_1$, $q_2$) is a duplicate pair. This is also called the time of duplicate link formation. The tag co-occurrence network is denoted as $G$ where nodes are the tags and edges have weights as already defined earlier.

\section{Duplicate question retrieval}

Suppose we have a set of questions $Q$ and graph $G$ (tag co-occurrence network). Given a pair of questions $(q_1, q_2)$, for $q_1$ (anchor question), our task is to find out its duplicate question $q_2$. In the subsequent sections, we will denote the anchor question as $q_a$, its actual duplicate question as $q^+$, and other questions which are not duplicates will be denoted by $q^-$. Given anchor question $q_a$, we intend to rank the possible duplicate questions according to decreasing order of duplicity scores. The position of the gold duplicate $ q^+$ can be found in this rank list and we evaluate the system's performance using Mean Reciprocal Rank (MRR) and recall rate at $k$ (RR@$k$). In subsequent paragraphs, we describe the strategy for sampling  $q^-$ and building the candidate set. \\

\begin{figure}[!ht]
%\vspace*{-.5cm}
\centering
\small
\begin{minipage}{.5\textwidth}
  \includegraphics[width=1.0\linewidth]{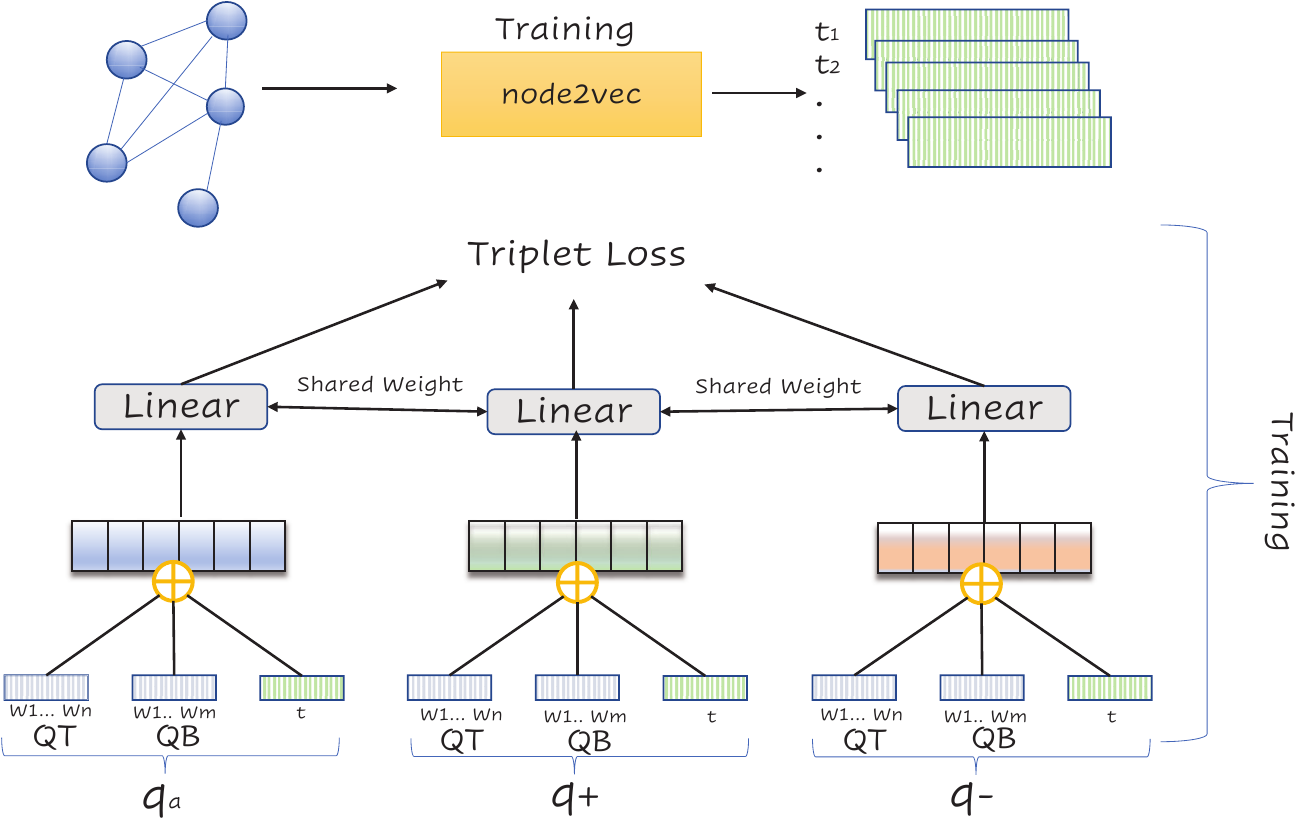}
    \caption{\footnotesize Training phase.}
  \label{fig:dupTrn}
\end{minipage}
\begin{minipage}{.49\textwidth}
 \centering
  \includegraphics[width=1\linewidth]{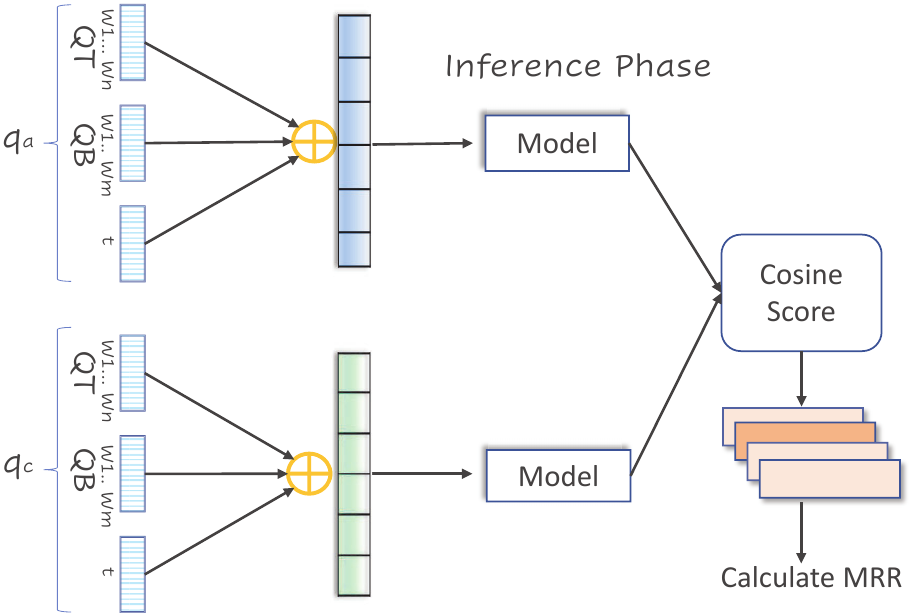}
    \caption{\footnotesize Inference phase.}
  \label{fig:dupTst}
\end{minipage}
%\vspace*{-.5cm}
\end{figure}
\noindent \textbf{Model architecture}: Our model consists of a transformer encoder with mean pooling for embedding generation followed by concatenation and linear transformation with activation. Going forward in this paper, we denote our method using \textbf{TE}: transformer encoder. We have used a Siamese neural network~\cite{Bromley:1993} for solving this problem. The pipeline of the proposed model is presented in Figure~\ref{fig:dupTrn}.\\
\textbf{Text embedding}: We use the representation of the title and body of the questions as features. As part of preprocessing these two pieces of text, we removed the URLs, stopwords, etc. We have used transformer encoders with mean pooling to generate the embeddings. For a given question $q$, the embeddings for the title and body are denoted as $e_{q}^\mathcal{QT}$ and $e_{q}^\mathcal{QB}$ respectively.\\
\textbf{Graph embedding}: We have used the tag co-occurrence network to compute tag features. These features are blended with the text to obtain the final embedding. We train the tag co-occurrence network using the node2vec~\cite{Grover:2016} algorithm. Here, we cannot train any graph neural network to obtain the embeddings of the node because we do not have any specific target variable\footnote{The unsupervised approach suitable for graph neural network also did not perform well.}. So, we did not go forward with this setup. We have ordered the tags based on their occurrence in training data. For every question, we have an ordered list of tags; from this list, we take only the embedding of the top tag ($e_q^{t}$) to be blended with the text.\\
%\textbf{Text and Graph embedding}: These are generated exactly as in case of duplicate question identification time prediction.\\
%\textbf{Graph embedding}: These are generated exactly as in case of duplicate question identification time prediction.\\ %We have used the tag co-occurrence network to compute tag features and associate them with the questions accordingly. For every question, we have an ordered list of tags, and we are considering embedding of the first tag that can be denoted by $e_q^{t}$.\\%\SB{Along with that, we train the tag co-occurrence network using the netMF~\cite{Qiu:2018} algorithm.}  \\
%We use the similar procedure of creating a tag co-occurrence network as mentioned in the previous experimental setup. we are considering embedding of the top tag that can be denoted by $e_q^{t}$. \\
\textbf{Concatenation and linear transformation}: In this step given a question $q$, we concatenate the feature vectors $e_{q}^\mathcal{QT}$ and $e_{q}^\mathcal{QB}$ for text-based model. For text+network based model, we concatenate the feature vectors $e_{q}^\mathcal{QT}$, $e_{q}^\mathcal{QB}$ and $e_q^{t}$. Concatenation is denoted by $\oplus$. 
% Thus $E_{q}^{TB} = [e_{q}^\mathcal{QT} \oplus e_{q}^\mathcal{QB}]$ and 	 
% $E_{q}^{\prime TB} = \sigma(W_{TB} \cdot E_{q}^{TB} + b_{TB})$.
\begin{equation} 
\small
\label{linear}
E_{q}^{TB} = [e_{q}^\mathcal{QT} \oplus e_{q}^\mathcal{QB}],
E_{q}^{\prime TB} = \sigma(W_{TB} \cdot E_{q}^{TB} + b_{TB})
\end{equation}

\noindent \textbf{Objective function}: Given an anchor question $q_a$, positive question $q^+$ and negative question $q^-$, the triplet margin loss\footnote{https://en.wikipedia.org/wiki/Triplet\_loss} tune the model ($\theta$) in such a way that the distance between $\theta(q_a)$ and $\theta(q^+)$ will decrease but the distance between $\theta(q_a)$ and $\theta(q^-)$ will increase. %We can use any distance metric preferably edit distance~\rh{Need change here}. 
We assume that we have got $E_{q_a}^{\prime TB}$, $E_{q^+}^{\prime TB}$, $E_{q^-}^{\prime TB}$ representation from the model $\theta$ for anchor question $q_a$, positive question $q^+$ and negative question $q^-$ respectively. The loss function is as follows - $L(\theta) =  max[d(E_{q_a}^{\prime TB}, E_{q^+}^{\prime TB}) - d(E_{q_a}^{\prime TB}, E_{q^-}^{\prime TB}), 0 ]$. Here the $\small d(E_{q_a}^{\prime TB}, E_{q^+}^{\prime TB}) = ||E_{q_a}^{\prime TB} - E_{q^+}^{\prime TB}||_p$. $p$ is the norm degree of the pairwise distance.
%\vspace{-0.6cm}
\begin{algorithm}[!ht]
%\scriptsize
\tiny
\caption{\label{algo:algo1} Negative sampling strategy}
\begin{algorithmic}
\State Buckets ${b_i}\in\bm{B}$, similarity matrix $\mathcal{B}\lbrack {b_1}, {b_2}\rbrack$ between the representation of the buckets $b_1$ and $b_2$; 
%\in \bm{|B|}\times\bm{|B|}$;
%\State There are positive pairs $(q_a, q^+)$ where $q_a, q^+ \in Q$;
\For{each positive pair $(q_a, q^+)$}
\State create an empty dictionary $D_{J, q_a}$;
\State get the bucket of $b_a \in \bm{B}$ to which $q_a$ belongs;
\State obtain an ordered list (high to low) of buckets $B_k \setminus b_a$ where the ordering is based on the similarity $\mathcal{B}\lbrack b_a, b_k \rbrack \wedge \mathcal{B}\lbrack b_a, b_k \rbrack >\alpha$;  %>\alpha \forall \bm{k} \in [0, \gamma)$; %Here, we are considering top 100 buckets. Replace |B| with certain theshold.
\For{each bucket $b_k \in B_k$}
\For{each $q \in b_k$}
\State calculate tag overlap $T_{overlap}(q_a, q)$;
\If{$q$ is answered }%\& $J(q_a, q) > \gamma$
\State append item $\{q: T_{overlap}(q_a, q)\}$ to the dictionary $D_{J, q_a}$;
\EndIf
\EndFor
\EndFor
\State choose the $q$ with maximum $T_{overlap}(q_a, q)$ from $D_{J, q_a}$;
\State $q \leftarrow q^-$;
\EndFor
\end{algorithmic}
\end{algorithm}
%\vspace{-0.5cm}

\noindent \textbf{Negative sampling strategy}: In the  training, we prepare triplets consisting of ($q_a$, $q^+$, $q^-$). This section discusses how we sample the $q^-$ for every duplicate pair present in the training data. We obtain buckets based on the duplicate clusters. Suppose there are four duplicate pairs ($q_1$, $q_2$), ($q_1$, $q_3$), ($q_3$, $q_4$) and ($q_5$. $q_6$). As the pairs ($q_1$, $q_2$), ($q_1$, $q_3$), ($q_3$, $q_4$) have transitive relationship, they would form a bucket whereas ($q_5$. $q_6$) is not transitive with them so it would form another bucket. We form the buckets out of all the duplicate pairs in the training set. We next obtain a representation of a bucket as average embeddings of all the questions in that bucket. Based on this representation, we compute the bucket-bucket similarity matrix $\mathcal{B}$. We then sample buckets most similar to the bucket containing $q_a$. Out of the questions in these similar buckets, we choose the one with the highest tag overlap with $q_a$ as $q^{-}$. The heuristic attempts to identify one of the hardest negative samples so that the decision boundary is robust. The detailed steps are given in Algorithm~\ref{algo:algo1}.

\noindent \textbf{Inference}: Given an anchor question, we obtain the similarity scores for the possible duplicate questions with the anchor and rank them during the inference. Before getting the scores, we must prepare a set of possible duplicate questions for a given anchor question. Given an anchor question $q_a$, we call this set a candidate set ($Q_a^c$). The detailed inference phase is presented in Figure~\ref{fig:dupTst}. This Figure shows the process of getting similarity scores between $q_a$ and $q_c \in Q_a^c$ from the model. The following section discusses the strategies used for generating the candidate set $Q_a^c$ for all the anchor questions.\\
\noindent \textbf{Candidate set generation}: We construct a candidate set $Q_a^c$ for each anchor question $q_a$ using the following selection heuristic - (i) the extent of tag similarity between anchor and candidate question, (ii) if the candidate question has an answer (either accepted or unaccepted), and (iii) the question title ($\mathcal{QT}$) similarity between the anchor and the candidate question. Our intuition is that most duplicate pairs have tags in common and a similarity in their question title.\\ %, and we observe a similar pattern in duplicate pairs~\footnote{https://askubuntu.com/questions/144237/how-do-i-uninstall-ubuntu-wubi}. In this pair of questions, they have the same tag called ``wubi''. 
%We calculate $Freq_{T}$ from the training segment, which contains the tag and its frequency in descending order. $Freq_{T}$ is a mapping contain $\{t_1:f_1, t_2: f_2, \dots t_k:f_k\}$.  
%Here, frequency $f_k$ denotes the number of questions where the same tag occurred. Given $q$, we sort the list of tags $T_q$ based on frequency in descending order. 
Let us denote the tag list of the anchor question as $T_{q_a}$. Further for each anchor question $q_a$, we create an empty candidate set $Q^{c}_{a}$. We collect the previously posted questions (strictly earlier to the anchor question), which have at least one of the tags common with $T_{q_a}$. Let us call this question set as $Q^{c}$ and the tag list of each $q^c \in Q^{c}$ as $T_{q^c}$. Next, we calculate the Jaccard overlap ($J(T_{q_a}, T_{q^c})$) between $T_{q_a}$ and $T_{q^c}$ for each $q^c$. We retain only those questions in $Q^{c}$ which have $J(T_{q_a}, T_{q^c})>0.15$. We further filter $Q^{c}$ to have only those questions that have been already answered. Finally, we proceed with the last filter retaining only those questions in $Q^{c}$ whose embeddings have a cosine similarity of 0.27 or more with the embedding of the anchor question. We populate $Q^{c}_{a}$ with the final set of questions present in $Q^{c}$.

\section{Duplicate confirmation time prediction}
%\vspace{-0.3cm}
Suppose we have a question and its textual data such as $\mathcal{QT}$, $\mathcal{QB}$. Our goal is to predict the time gap $ts_{Gap}$ between the time when the recent most question in the pair is reported as a duplicate and the duplicate link formation time (i.e., when the question is closed as a duplicate). Thus, given a question pair ($q_1$, $q_2$), we have computed the time gap $ts_{Gap}(q_1, q_2)$ as ($ts(q_1, q_2) - ts(q_1)$) where $ts(q_1) > ts(q_2)$. We want to predict the time gap $ts_{Gap}$ for all the possible duplicate pairs. We sort the $|ts_{Gap}|$ in descending order and focus on the pairs that took a long time to close as duplicates. The idea is to present the top-ranked pairs with a long time gap to the moderators so that it could be addressed quickly. \\
%writing motivation of this problem
Note that our assumption here is that the recently posted question has already been marked as a duplicate of some earlier question by some user (regular user/moderator) but has yet to receive the necessary attention~\footnote{https://askubuntu.com/help/duplicates} from the moderators or anyone having more than 3K reputation. Unless the recently posted question gets a certain number of votes (usually takes a minimum of 5 votes) from the moderators/experienced users, the question is not considered a duplicate of the earlier question (i.e., the question link formation cannot take place). Our idea is to early predict those pairs which have remained ``open'' for a long (i.e., $ts_{Gap}$ is large) and facilitate their closing by bringing them to the notice of the moderators. We have the gold $ts(q_1, q_2)$ during the evaluation, and we compare the gold ranks with the predicted ranks using rank correlation methods.\\
\textbf{Text and graph embedding}: These are generated exactly the same as in case of duplicate question retrieval.\\
\noindent \textbf{Model architecture}: We use two models for this task -- (i) Decision Tree (DT), (ii) XGBoost (XGB) and (iii) Multi-layer perceptron (MLP).\\
We use the standard DT regressor~\cite{Breiman:1983} and XGBoost regressor~\cite{10.1145/2939672.2939785} with inputs as (i) text and (ii) text+ network features. The output is a regression score with duplicate pairs requiring the largest time to close. %Also, we use the existing XGBoostregressor to predict the time with inputs as text and text+network features.

In case of MLP, we use L1 loss ~\footnote{https://pytorch.org/docs/stable/generated\\/torch.nn.L1Loss.html} between the predicted time gap ($ts_{Gap}^{\prime}$) and the gold time gap ($ts_{Gap}$). The model architecture is summarized by equations ~\ref{Itime:linear12},~\ref{Itime:linear34} and ~\ref{Itime:linear5}. Given a pair of questions $(q_1,q_2)$, we use the same model for both features.
\begin{equation}
\small
\label{Itime:linear12}
\begin{aligned}
E_{q_1}^{TB} = ReLU(W_1^{TB} \cdot [e_{q_1}^{\mathcal{QT}} \oplus e_{q_1}^{\mathcal{QB}}] + b_1^{TB})\\
E_{q_2}^{TB} = ReLU(W_2^{TB} \cdot [e_{q_2}^{\mathcal{QT}} \oplus e_{q_2}^{\mathcal{QB}} ] + b_2^{TB})
\end{aligned}
\end{equation}
\begin{equation}
\small
\label{Itime:linear34}
\begin{aligned}
E_{q_1}^{\prime TB} = ReLU(W_1^{\prime TB} \cdot E_{q_1}^{TB} + b_1^{\prime TB})\\
E_{q_2}^{\prime TB} = ReLU(W_2^{\prime TB} \cdot E_{q_2}^{TB} + b_2^{\prime TB})
\end{aligned}
\end{equation}
\begin{equation}
\small
\label{Itime:linear5}
\begin{aligned}
%E_{q_12}^{\prime\prime TB}
ts_{Gap}^{\prime} = TanhShrink(W_{12}^{\prime\prime TB} \cdot [E_{q_1}^{\prime TB} \oplus E_{q_2}^{\prime TB}] + b_{12}^{\prime\prime TB})\\
%E_{q_2}^{\prime TB} = \sigma(W_2^{\prime TB} \cdot E_{q_2}^{TB} + b_2^{\prime TB})
\end{aligned}
\end{equation} 
%  Thus, $E_{q_1}^{TB} = ReLU(W_1^{TB} \cdot [e_{q_1}^{\mathcal{QT}} \oplus e_{q_1}^{\mathcal{QB}}] + b_1^{TB})$,
% $E_{q_2}^{TB} = ReLU(W_2^{TB} \cdot [e_{q_2}^{\mathcal{QT}} \oplus e_{q_2}^{\mathcal{QB}} ] + b_2^{TB})$, $E_{q_1}^{\prime TB} = ReLU(W_1^{\prime TB} \cdot E_{q_1}^{TB} + b_1^{\prime TB})$, and 
% $E_{q_2}^{\prime TB} = ReLU(W_2^{\prime TB} \cdot E_{q_2}^{TB} + b_2^{\prime TB})$. Finally, $ts_{Gap}^{\prime} = TanhShrink(W_{12}^{\prime\prime TB} \cdot [E_{q_1}^{\prime TB} \oplus E_{q_2}^{\prime TB}] + b_{12}^{\prime\prime TB})$. 

In the case of text+network features, we change the input embedding, i.e., instead of $[e_{q_1}^{\mathcal{QT}}, \oplus e_{q_1}^{\mathcal{QB}}]$, we pass $[e_{q_1}^{\mathcal{QT}} \oplus e_{q_1}^{\mathcal{QB}} \oplus e_{q_1}^{t}]$ as an input. $W_1^{TB}$, $W_2^{TB}$, $W_1^{\prime TB}$, $W_2^{\prime TB}$, $W_{12}^{\prime\prime TB}$ are the trainable weights. $b_1^{TB}$, $b_2^{TB}$, $b_1^{\prime TB}$, $b_2^{\prime TB}$ and $b_{12}^{\prime\prime TB}$ are the trainable biases. In the last layer, we use TanhShink\footnote{https://pytorch.org/docs/stable/generated/\\torch.nn.Tanhshrink.html} because of the span of the data where the lowest target is negative and the highest target value is positive.

\section{Experiments and results}
%~\rh{as the results reported in percentage, we have to provide \% in the result tables}
\subsection{Duplicate question retrieval}\label{dupret}
\noindent \textbf{Upper bound}: To calculate the upper bound\footnote{This term is adapted from~\cite{HazraAGMC21}}, given an anchor question, we identified whether the actual duplicate is present in the candidate set or not. If it is present in the candidate set, we consider the rank 1; otherwise, 0. For the test data, we obtained an upper bound of 62.8\%. Thus this is the best possible recall that can be achieved.

%\noindent \textbf{Baselines}: We use various baselines to compare our models.

\noindent\textit{Text features}: We use InferSent~\cite{Conneau:2017}%\footnote{InferSent version-1, word embedding dimension = 300, LSTM dimension = 2048, \textit{pooling} = `max').}
, BM25~\cite{Robertson:2009}%\footnote{Best values for parameter 'b' and 'k1' are 0.75 and 1.5 respectively}
,  Glove + BiLSTM~\cite{Pennington:2014,Hochreiter:1997}%\footnote{learning rate = 1e-5, batch size = 64, \#words = 256, respectively, BiLSTM hidden dimension \& final output dimension = 64.}
, word2vec + BiLSTM~\cite{Wang:2020}%\footnote{learning rate = 1e-4, batch size = 64, BiLSTM hidden dimension \& final output dimension = 64.}
, word2vec~\cite{Mikolov:2013} (word2vec algorithm directly trained on our CQA corpus%\footnote{window size=10, \#epochs = 5, learning rate = 0.025, final representation = averaged over all words, MLP layer (Figure~\ref{fig:dupTrn}) parameters: \#epochs is 40, $\epsilon$ = 1e-8, learning rate = 1e-3, output dimension = 512 (2048 for \textit{word2vec+network}).}
) and doc2vec~\cite{Le:2014} (doc2vec algorithm directly trained on our CQA corpus%\footnote{All parameters same as word2vec except for the output dimension in the MLP layer which is = 2048 (1024 for \textit{word2vec+network}.}
) to generate text embeddings. All the hyperparameters used in these baselines are obtained through grid search and are noted in Table~\ref{tab:hyperparameters}.

\noindent\textit{Network features}: While both the word2vec and doc2vec models discussed above are text only, here we add the network features obtained from the node2vec embeddings. The title, body, and tag embeddings are fed to the MLP layer (Figure~\ref{fig:dupTrn}).

\noindent \textbf{Experimental setup for our method (TE)}: We divide all the questions into three parts -- training, validation, and test. In training, we consider the duplicate pairs closed between 2010 to 2018, whereas, for the validation set, we use the last three months' data from 2019. For testing, we use the last three months' data from 2020. Since we follow a retrieval-like evaluation, we need to compare anchor questions with every candidate question in the candidate set. Thus the total number of comparisons being relatively high, we have chosen only three months of data for validation and testing.
We have a total number of $\sim$ 32K positive pairs in the training set. In the inference phase, based on the candidate set generation heuristic, the average number of questions in a candidate set is 5941\footnote{Without the candidate set generation strategy, the number of earlier questions to which the anchor question would have to be compared would be close to $\sim$ 300K.} for test data. In our test data, we have 485 anchors, thus making the total number of comparisons equal to almost 2.8 million ($485 \times 5941$). %In the validation set, we have a total number of 641 anchors.\\

\noindent\textit{Specifications of the text embedding generation}: We have used {\em multi-qa-MiniLM-L6-cos-v1}\footnote{https://huggingface.co/sentence-transformers/multi-qa-MiniLM-L6-cos-v1} pretrained model to generate the embeddings of the $\mathcal{QT}$ and $\mathcal{QB}$. The default embedding dimension is 384. This model uses the pretrained setup with 6 layer version of Microsoft/MiniLM-L12-H384-uncased by keeping only every second layer\footnote{https://huggingface.co/nreimers/MiniLM-L6-H384-uncased}. %Further, they finedtuned the model by combining multiple datasets.\\

\noindent\textit{Specifications of the network embedding generation}: We investigate different values of the parameters for training node2vec through grid search and populate  64-dimensional embedding. We got $p$ as 1.3, $q$ as 0.8, the number of the walk as 5, the walk length as 80, $min\_count$ as 3, $batch\_word$ as 5, and parameter $window$ to 10.\\ %We generate 64-dimensional embedding for each node.
%This model uses the pre-trained model of 6 layer version of Microsoft/MiniLM-L12-H384-uncased by keeping only every second layer~\footnote{https://huggingface.co/nreimers/MiniLM-L6-H384-uncased}. Further, they fine-tuned the model by combining multiple datasets.\\
 %as well as the NetMF~\cite{Qiu:2018}. Here, we cannot use a graph neural network due to the lack of proper labels of the node in the network. \\
 \noindent \textit{Hyperparameters}: For the hyperparameter tuning of the text-only models, we have found the learning rate as 1e-3 and $\epsilon$ as 1e-8. The output size of the representation is 512 and, the number of epochs is 40.\\ %we  different values of the parameters. After grid search, we set the $p$ to 1.3 and $q$ to 0.8. The number of the walk is set to 5, and the walk length is kept as 80. The $min\_count$ is set to 3 and $batch\_word$ is set to 5. The parameter $window$ is 10. Using node2vec, we are generating the 64-dimensional embedding for each node.
%For the NetMF algorithm, we have tuned the hyperparameter as follows. We set $ord$ (number of PMI matrix powers) to 2, and the iteration is set to 10. The number of negative samples is set to 5, and the embedding dimension is set to 16. So, we have to consider node2vec based features for all the baselines. 
\noindent \textit{Evaluation metrics}: We use the mean reciprocal rank (MRR) and recall rate (RR@$k$) to evaluate all the models. We have used different values of $k$ ranging from 10 -- 500. Since the candidate set size is $\sim$ 5K, RR@$500$ is expected to present good suggestions to the moderators for duplicate question closure, reducing the otherwise tremendous manual load. Note that the evaluation results presented here are only for those anchor questions that have duplicates.\\ %For evaluation, we have only considered the only those anchor questions which have duplicate. We try different values of k such as 10, 20, 30, 50, 100 and 500. As the size of the average candidate set is quite high (5.9K), RR@500 will be able to provide the necessary support to the moderators/experienced users to find the duplicates efficiently.\\

\noindent \textbf{Baselines}: We use nine different baseline methods. 
\begin{table*}
%\vspace*{-0.6cm}
%\begin{minipage}[c]{0.45\textwidth}
\centering
%\scriptsize
\tiny
\resizebox{.99\textwidth}{!}{
\begin{tabular}{|c|c|c|c|c|c|c|c|} \hline %p{4cm}
%\multirow{2}{*}{\bf Training data} & \multicolumn{3}{c|}{\bf Text based}\\
%\cline{2-4}
$q_a$ & {Title of $q_a$} & $q^+$ & {Title of $q^+$} & PR & PR & N & \#w(T) \\%& \#w(B)
& & & & (TE) & (TE+net) & & \\
  \hline
%\cline{1-3}
 %{\bf Glove}+regressor &  &  & \\ \hline
 359751 & {Prefix argument for starting chromium with hardware acceleration} & 128126 & {How to execute a command with "=" sign in a desktop shortcut?} & 55 & 89 & 28 & 8  \\ \hline %& 283
 364255 & {Running, or "injecting" software with specific date} & 250575 & {Change Ubuntu time and date for specific application} & 2 & 3 & 16  & 10 \\ \hline %& 36
 363710 & {How to change Ubuntu 20.04 Desktop file manager (not gnome)?} &  338041 & {How to remove GNOME Shell from Ubuntu 20.04 LTS to install other desktop environment from scratch?} & 62 & 147 & 26	& 13  \\ \hline %& 59
 364236 & {how I would make Ubuntu GUI in wsl subsystem in Window} & 262015 & {What's the easiest way to run GUI apps on Windows Subsystem for Linux as of 2018?} & 4 & 10 & 31	& 11 \\ \hline %& 46
364973 & {Why do I have to use sudo if I am the only user?} & 245098 & {What's exactly the point of the sudo command, in terms of security?} & 8 & 13 & 36 & 14 \\ \hline %&  75
%1802821	& 8.338 & 5.625 & 7.860 &  2.712 & 0.477 & 26\\ \hline
\end{tabular}
}
%\caption{Test examples where the \textbf{TE} model predicts better ranks of actual duplicate question than the \textbf{TE+network} model.}
%\end{minipage}
\end{table*}
\begin{table*}
%\begin{minipage}[c]{0.45\textwidth}
\centering
%\scriptsize
%\tiny
\tiny
\resizebox{.99\textwidth}{!}{
\begin{tabular}{|c|c|c|c|c|c|c|c|} \hline %p{4cm}
%\multirow{2}{*}{\bf Training data} & \multicolumn{3}{c|}{\bf Text based}\\
%\cline{2-4}
$q_a$ & {Title of $q_a$} & $q^+$ & {Title of $q^+$} & PR & PR & N & \#w(T) \\ %& \#w(B)
& & & & (TE) & (TE+net) & & \\
  \hline
%\cline{1-3}
 %{\bf Glove}+regressor &  &  & \\ \hline
363584 & {t440 ubuntu drivers} & 140121 & {How to download all required Ubuntu drivers} & 45 & 39	& 89 & 3  \\ \hline %& 46
362424 & {Cannot Update packages} & 3491 & {How do I fix the GPG error "NO\_PUBKEY"?} & 4868 & 1363 & 49  & 4 \\ \hline %& 52
363917 & {Remove plasma from ubuntu} & 187651  & {How to remove KDE Plasma-Desktop?} & 4  & 3 & 68	& 4 \\ \hline %& 33
359502 & {Ubuntu Live USB boot problem} & 45554 & {My computer boots to a black screen, what options do I have to fix it?} & 3621 & 59 & 68	& 5 \\ \hline %& 63
%363620 & 127967 & 2967.5 & 2640 & 73 & 5 & 103 \\ \hline
365800 & {AMD drivers Ubuntu 20.04.1} & 210683 & {Ubuntu 14.04.5/16.04 and newer on AMD graphics} & 80 & 43 & 54 & 4 \\ \hline %& 84 
%1802821	& 8.338 & 5.625 & 7.860 &  2.712 & 0.477 & 26\\ \hline
\end{tabular}
}
%\end{minipage}
\caption{\label{tab:erroranalysis} \footnotesize $q_a$: anchor question, $q^+$: actual duplicate, PR(TE): Predicted rank of \textbf{TE}, PR(TE+net): Predicted rank of \textbf{TE+network}, \textbf{N}: \#neighbors, \#w(T): \#words in title. Test examples where (up) the \textbf{TE} model predicts better ranks of actual duplicate question than the \textbf{TE+network} model, (down) \textbf{TE+network} model predicts better ranks of actual duplicate question than \textbf{TE} model.%\#w(B): \#words in body
}
%\vspace*{-1.5cm}
\end{table*}
\noindent \textbf{InferSent}~\cite{Conneau:2017}: Infersent is a sentence encoder where the representation of each sentence has been computed. It is a BiLSTM network with max pooling. We compute embeddings for each question and subsequently obtain the cosine similarity between the anchor question and its candidates. \\
\noindent \textbf{BM25 Search}~\cite{Robertson:2009}: We use the standard unsupervised method for BM25 search. Here, we provide the title and body of the questions to train the BM25 model. Further, for each query, we obtain scores of its candidates and rank the candidates based on the scores (higher score corresponding to better rank).\\
\noindent \textbf{Glove + BiLSTM}~\cite{Pennington:2014, Hochreiter:1997}: For each question, we extract the 300d representation of words and further use BiLSTM and one linear layer to obtain the final representation. Given a triplet of questions during the training, we compute the triplet loss between anchor, positive and negative questions. During the evaluation, we use cosine similarity.\\
\noindent \textbf{word2vec + BiLSTM}~\cite{Wang:2020}: In this case, we use the pretrained `Google News-vectors-negative300' model to obtain the embedding of the words present in the texts of each question. Due to the Siamese-like architecture, we do not use the sigmoid activation mentioned in their architecture. The rest of the architecture is verbatim similar. During training, we feed the embedding of a sequence of words to BiLSTM and one linear layer to get the final fixed-length representation for a question. Here, we use triplet loss. During the evaluation, we use cosine similarity to rank the candidates.  \\
\noindent \textbf{word2vec}~\cite{Mikolov:2013}: Everything else remaining the same as the \textbf{TE} model, we replace the transformer encoders with word2vec to generate the question title and body embeddings. The word2vec embeddings are obtained by training the word2vec algorithm from scratch using the entire CQA corpus. All parameters were identified through a grid search.\\
\noindent \textbf{doc2vec}~\cite{Le:2014}: We replace the transformer encoder with the trained doc2vec to generate question titles and body embeddings. We train the doc2vec model using the whole CQA corpus. Further, we obtain the question title and body embeddings from the trained doc2vec models.\\
\noindent \textbf{DupPredictor}~\cite{Zhang:2015}: We implement DupPredictor algorithm and test it on our dataset. We create four components -- title similarity component, question body similarity component, topic similarity component, and tag similarity component. For topic modeling, we train the LDA model on the whole corpus (concatenating the title and the body of a question). The number of topics is 100.\\
\noindent \textbf{Dupe}~\cite{Ahasanuzzaman:2016}: We implement the Dupe method for our dataset. In their paper, they concluded that title, body, tag, title-body, body-title, title-tag, code similarity features are contributing to the best performance. So, compute these features on our dataset. We use the logistic regression as mentioned in the paper. \\
\noindent \textbf{SBERT STSb models}~\cite{reimers-2019-sentence-bert}: We use two pretrained models for obtaining the title and the body embedding of the question. The pretrained models are distilbert-base-nli-stsb-quora-ranking and distilbert-multilingual-nli-stsb-quora-ranking. Further, we feed the embedding to our MLP model.

To compare our model with the existing methods, we treat the models in the Siamese network setup. All the hyperparameters used in these baselines are obtained through grid search and are noted in Table~\ref{tab:hyperparameters}. %are given in appendix~\ref{sec:appendix}.\\
\begin{table*}[!ht]
%\vspace*{-.5cm}
%\small
%\footnotesize
\centering
\resizebox{.75\textwidth}{!}{
\begin{tabular}{|p{2.5cm}|p{1.2cm}|p{1.2cm}|p{1.2cm}|p{1.2cm}|p{1.2cm}|p{1.2cm}|p{1.2cm}|} \hline
{\bf Methods} & {\bf MRR} & {\bf RR@10} & {\bf RR@20} & {\bf RR@30} & {\bf RR@50} & {\bf RR@100} & {\bf RR@500}\\ \hline
\multicolumn{8}{|c|}{\bf Text only} \\ \hline

%\cline{2-13}
{\bf word2vec} & 4.980  & 7.628 & 10.309 & 13.814 & 17.319 & 21.855 & \underline{39.175}\\ \hline
{\bf doc2vec} & 0.840 & 1.440 & 2.061 & 2.061 & 4.536 & 9.278 & 23.505\\ \hline
{\bf Pretrained word2vec + BiLSTM}&  { 2.299} & { 3.505} & { 4.123} & { 5.154} & { 6.391} & { 8.247} & { 18.556}\\ \hline
{\bf Glove + BiLSTM}&  { 1.403} & { 2.474} & { 3.711} & { 4.123} & { 4.742} & { 6.597} & { 15.463}\\ \hline
{\bf BM25 Search}&  {\underline{6.060} } & {\underline{10.300}} & { \underline{13.190}} & {14.840} & {\underline{17.730}} & {\underline{24.740}} & { 37.930}\\ \hline
{\bf InferSent}&  { 3.200} & { 4.120} & { 6.180} & { 7.210} & { 8.650} & { 11.340} & { 22.680}\\ \hline
{\bf DupPredictor} & {4.560} & {10.100} & {12.780} & {\underline{15.250}} & {17.310} & {21.850} & {35.870} \\ \hline
{\bf DUPE} & {2.750} & {3.910} & {5.360} & {7.210} & {9.480} & {12.780} & {24.740}  \\ \hline
{\bf TE}&  {\bf 9.452} & {\bf 15.876} & {\bf 19.381} & {\bf 21.649} & {\bf 25.154} & {\bf 31.546} & {\bf 44.948}\\ \hline
%{\bf Multiple Negative Ranking loss} & & & & & &  &\\ \hline
%{\bf ASIM}& & & & & &  &\\ \hline\hline
\multicolumn{8}{|c|}{\bf Text+network} \\ \hline

{\bf word2vec + network} & \underline{4.980} & 8.453 & 12.371 & 14.226 & 17.113 & 22.680 & \underline{40.618}\\ \hline
{\bf doc2vec + network} & 0.740 & 1.649 & 2.886 & 3.298 & 5.154 & 8.041 & 23.711\\ \hline
{\bf SBERT STSb distillbert + network} & 4.190 & \underline{10.220} & \underline{12.710} & \underline{15.080} & \underline{18.100} & \underline{24.200} & 38.770 \\ \hline
{\bf SBERT STSb distillbert multilingual + network} & 3.490 & 7.180 & 10.820 & 13.190 & 15.400 & 19.880 & 33.000 \\ \hline
{\bf TE+network} & {\bf11.088*} & {\bf 18.350} & {\bf 23.917} & {\bf 27.010} & {\bf 32.164} & {\bf 36.082} & {\bf 46.597} \\ \hline
%{\bf ASIM + network} & & & & & &  &\\ \hline
%{\bf Google + network} & & & & & &  &\\ \hline
\end{tabular}
}
\caption{\footnotesize Duplicate question retrieval. All the results are shown in percentages. TE: transformer encoder, *: the result of the text+network model is significantly different ($p<0.03$ using M-W U test) from the text-only model.}
\label{tab:dup_results}
%\vspace*{-0.6cm}
\end{table*}
%\vspace{-0.5cm}

\begin{table}[h]
%\begin{minipage}[c]{0.5\textwidth}
\tiny
\centering
\resizebox{.75\textwidth}{!}{\begin{tabular} {|p{1.5cm}|p{5.8cm}|}\hline 
\textbf{Method} & \textbf{Hyperparameters} \\ \hline
InferSent & version-1, word embedding dimension = 300, LSTM dimension = 2048, pooling = `max'. \\ \hline
BM25 & $b=0.75$ and $k1=1.5$ \\ \hline
Glove + BiLSTM & learning rate = 1e-5, batch size = 64, \#words = 256, BiLSTM hidden dimension \& final output dimension = 64. \\ \hline
word2vec + BiLSTM & learning rate = 1e-4, batch size = 64, BiLSTM hidden dimension \& final output dimension = 64. \\ \hline
word2vec & window size = 10, \#epochs = 5, learning rate = 0.025, final representation = averaged over all words, MLP layer (Figure~\ref{fig:dupTrn}) parameters: \#epochs = 40, $\epsilon$ = 1e-8, learning rate = 1e-3, output dimension = 512 (2048 for \textit{word2vec+network}). \\ \hline
doc2vec & All parameters same as word2vec except for the output dimension in the MLP layer = 2048 (1024 for \textit{word2vec+network}.) \\ \hline
DupPredictor & $\alpha$ = 0.8, $\beta$ = 0.8, $\gamma$ = 0.1 and $\delta$ = 0.6\\ \hline
Dupe & batch size = 32, learning rate = 1.00E-05, number of epochs = 40\\ \hline
SBERT STSb distillbert base  & \#epochs = 30, $\epsilon$ = 1e-7, learning rate = 1e-4  \\ \hline
SBERT STSb distillbert multilingual &  \#epochs = 20, $\epsilon$ = 1e-7, learning rate = 1e-4\\ \hline
\end{tabular}}
\caption{\label{tab:hyperparameters}\footnotesize Hyperparameters for the baseline methods chosen based on grid search.}
%\end{minipage}
%\vspace{-1.15cm}
\end{table}

%We make a few observations from the results presented in Table~\ref{tab:dup_results}. Our method based on a transformer encoder outperforms all the ways in text-based settings. Also, the transformer encoder performs better than the models based on word2vec and doc2vec, both in text+network settings. The inclusion of network features brings in statistically significant improvements. The RR@$500$ in the TE+network settings reaches 46.5\%, closest to the upper bound, leaving future scopes for further improvement.
\noindent\textbf{Results}: We make a few observations from the results presented in Table~\ref{tab:dup_results}. Our method based on a transformer encoder (\textbf{TE}) outperforms all the other approaches in text-based settings. We present the results for MRR and RR@\{10, 20, 30, 50, 100, 500\}. The table shows that our proposed technique performs better than all baselines, including state-of-the-art \textbf{DUPE} and \textbf{DupPredictor}. Also, we observe exciting performance improvement in several other popular baselines. For example, \textbf{BM25 search} achieves an MRR score of 6.06 whereas our method achieves an MRR score of 9.452 (an increase of almost 3.5\%). Similarly, we also observe an increase of nearly 5\%, 6\%, 8\%, and 7\% of RR values at RR@10, RR@20, RR@50, and RR@50 respectively, for our method (\textbf{TE}) over \textbf{BM25 search}, which secures a second position across the different baselines. At RR@30 \textbf{DupPredictor} performs better than \textbf{BM25 search}, but our method outperforms \textbf{DupPredictor} by almost 6\%. We note an improvement for the case of word2vec at RR@500 when compared to all other baselines; however, our technique outperforms word2vec by over 5\%. So, with all types of evaluation metrics, our proposed method (\textbf{TE}) consistently achieves better results than all other baselines.\\
We also observe a similar trend in the \textbf{text+network} model. Here we see an increase in MRR score when compared with \textbf{word2vec + network}. Similarly we achieve 18.35\% in RR@10, 23.917\% in RR@20, 15.08\% in RR@30, 18.1\% in RR@50 and 36.082\% in RR@100 which outperforms its nearest competitor \textbf{SBERT STSb distilbert + network} with increase of almost 8\% in RR@10, 11\% in RR@20, 12\% in RR@30, 14\% in RR@50 and 12\% in RR@100 respectively. We observe our proposed method \textbf{TE+network} reaches 46.597\% for RR@500, which outperforms \textbf{word2vec+network} with a margin of almost 6\%.
%\am{A lot of mixing of past and present tense. Either keep everything as present tense or everything as past tense.}
\subsection{Duplicate confirmation time prediction}
%\vspace{-0.45cm}
\label{duptime}
\noindent \textbf{Experimental setup}: We have considered all the duplicate pairs present in the dataset in this setup. Further, we divide the dataset into train, validation, and test sets. For training, we have considered pairs of questions where all the questions were posted before 2020. Further, we use 25\% of this training set for validation. For testing, we have considered all the pairs where the questions were posted after 2020. We have considered the time gap in hours. In specific, we predict $log_{10}(ts_{Gap})$ using two models -- (i) Decision Tree (DT) and (ii) XGBoost (XGB) (iii) Multilayer perceptron (MLP).\\
Here again, we have used {\em multi-qa-MiniLM-L6-cos-v1}%\footnote{https://huggingface.co/sentence-transformers/multi-qa-MiniLM-L6-cos-v1} 
 pre-trained model to generate the embeddings of $\mathcal{QT}$ and $\mathcal{QB}$. The embedding dimensions for each of them are 384. To get the node embeddings from the tag co-occurrence network, we have used the node2vec~\cite{Grover:2016} algorithm. For training, the same parameters noted in section~\ref{dupret} have been used.\\
\textit{Settings for the DT model}: For the DT model, after the parameter tuning, the criterion is set to squared error, splitter is set to `best', max-depth is set to 7, and min\_samples\_split is set to 2. For the text only model, we concatenate the title and the body embeddings of a question $q_i$ to obtain a 768 dimensional embedding -- $[e_{q_i}^{\mathcal{QT}} \oplus e_{q_i}^{\mathcal{QB}}]$. For a pair $(q_1, q_2)$ we feed the DT with $[e_{q_1}^{\mathcal{QT}} \oplus e_{q_1}^{\mathcal{QB}} \oplus e_{q_2}^{\mathcal{QT}} \oplus e_{q_2}^{\mathcal{QB}}]$ as the feature. For the text+network model we feed the DT with $[e_{q_1}^{\mathcal{QT}} \oplus e_{q_1}^{\mathcal{QB}} \oplus e_{q_1}^{t} \oplus e_{q_2}^{\mathcal{QT}} \oplus e_{q_2}^{\mathcal{QB}} \oplus e_{q_2}^{t}]$ as the feature where $e_{q_i}^{t}$ represents the 64 dimensional embedding of the top tag of $q_i$ obtained from the tag co-occurrence network.\\%For a pair $(q_1, q_2)$ as an instance, we provide input as $[e_{q_1}^{\mathcal{QT}} \oplus e_{q_1}^{\mathcal{QB}} \oplus e_{q_1}^{t}]$ for $q_1$ and $[e_{q_2}^{\mathcal{QT}} \oplus e_{q_2}^{\mathcal{QB}} \oplus e_{q_2}^{t}]$ for $q_2$ to the model.}\\ %The experimental results are given in Table~\ref{tab:main_results}. \\
\textit{Settings for XGB model}: We have used the same setup as the DT model for text and text+network-based models. After tuning the parameters, the \textit{n estimator, max depth} are kept as 1000 and 7, respectively. The value of \textit{eta, subsample and colsample\_bytree} are set to 0.1, 0.7 and 0.8 respectively.\\
\textit{Settings for the MLP model}: As in the DT setting, here also we use the same $[e_{q_i}^{\mathcal{QT}} \oplus e_{q_i}^{\mathcal{QB}}]$ embedding to represent a question in the text-only model. For a given a pair $(q_1, q_2)$, we pass the corresponding 768 dimensional representations of $q_1$ and $q_1$ to the input layer. The intermediate hidden layer is set to 256 and 64. The final output layer size is set to 1. We found the batch size is 64 and the learning rate is 2e-5.
For the text+network model, each question is again an 832 (384+384+64) dimensional embedding of the form $[e_{q_i}^{\mathcal{QT}} \oplus e_{q_i}^{\mathcal{QB}} \oplus e_{q_i}^{t}]$. The corresponding 832 dimensional representations of $q_1$ and $q_2$ for the pair $(q_1, q_2)$ are fed to the input layer of the MLP. The intermediate hidden layers are set to 512 and 64. The identified batch size is 64 and learning rate is 2e-5.

\begin{table}[!htb]
\scriptsize
\centering
\tiny
\resizebox{.75\textwidth}{!}{\begin{tabular}{|c|c|c|c|c|c|} \hline %p{4cm}
%\multirow{2}{*}{\bf Training data} & \multicolumn{3}{c|}{\bf Text based}\\
%\cline{2-4}
$q_a$  & $q^+$ & PR (Dup- & PR & PR & PR \\ %& \#w(B)
 & & Predictor) & (DUPE) & (TE) & (TE+net)  \\
  \hline
%\cline{1-3}
 %{\bf Glove}+regressor &  &  & \\ \hline
363584  & 140121 & {24} & {308} & {45}	& {39}   \\ \hline %& 46
362424   & 3491 & {3424} & {3324} & {4868} & {1363} \\ \hline %& 52
363917   & 187651  & {579} &  {2655} & {4} & {3} \\ \hline %& 33
359502   & 45554 & {2227} & {4144} & {3621} & {59}\\ \hline %& 63
%363620 & 127967 & 2967.5 & 2640 & 73 & 5 & 103 \\ \hline
365800 & 210683 & {17} & {2607} & {80} & {43} \\ \hline %& 84 
%1802821	& 8.338 & 5.625 & 7.860 &  2.712 & 0.477 & 26\\ \hline
\end{tabular}}
\caption{\label{tab:comparison} \footnotesize $q_a$: anchor question, $q^+$: actual duplicate, PR (DupPredictor): predicted rank of \textbf{DupPredictor}, PR (DUPE): predicted rank of \textbf{DUPE}, PR (TE): predicted rank of \textbf{TE}, PR (TE+net): predicted rank of \textbf{TE+network}}
\end{table}

\begin{table}[!htb]
%\vspace{-0.2cm}
%\scriptsize
\tiny
\centering
%\begin{minipage}[c]{0.42\textwidth}
\resizebox{.75\textwidth}{!}{\begin{tabular}{|p{2.8cm}|p{1cm}|p{1cm}|} \hline %|p{2cm}|p{2cm}
%\centering
{\bf Methods} & {\bf RMSE} & {\bf $\rho$} \\ \hline %& {\bf $\tau$@10} & {\bf $\rho$@10}
{\bf Text-DT} & 1.336   & 0.130 \\ \hline
{\bf Text-XGB} & 1.278 & 0.189 \\ \hline
{\bf Text-MLP} & 1.186   & 0.208 \\ \hline \hline %& 0.288 &  0.418
%{\bf SVR} & 1.184 & 0.39999 & 0.6 & 0.422 & 0.5757 & 0.1563 & 0.2329\\ \hline \hline
{\bf Text+Network-DT} & 1.312 & 0.130*\\ \hline
{\bf Text+Network-XGB} & 1.262 & 0.202* \\ \hline
{\bf Text+Network-MLP}  & \textbf{1.180}  & \textbf{0.213}* \\ \hline %0.19999 & 0.27272 &
\end{tabular}}
\caption{\label{tab:main_results}\footnotesize Duplicate question confirmation time prediction. $\rho$: Spearman's rank correlation, *: Results of text+network models are significantly different from the text only models with $p<0.01$ using M-W U test.}
%\end{minipage}
\end{table}

\noindent\textbf{Results}: The experimental results are presented in Table~\ref{tab:main_results}. The least RMSE is obtained for the text+network model using MLP. The Spearman's rank correlation ($\rho$) between the gold and the predicted rank for all the 3756 test pairs is also best for the text+network model using MLP. Given such a massive list of pairs, we believe that our results are pretty impressive. Further, we observe that adding network features always brings statistically significant improvements.
%MLP text model: 5 linear layer, 2-2 -> 768, 256, 64| 768, 256, 64 | 64*2,1 | 2-2 relu | 1 tanhs | lr = 2e-5 | bacths size= 64 | log base 10
%text+ graph: 5 linear layer, 2-2 -> 832, 512, 64| 832, 512, 64 | 64*2,1 | 2-2 relu | 1 tanhs | lr = 2e-5 | bacths size= 64 | log base 10
%\subsection{Observations}
%\noindent\textbf{Duplicate question identification time:} In this problem, we have tried two models on the text-based features and text+network-based features -- (i) decision tree (DT) (ii) XGBoost and (iii) MLP. For the text-based model, we observe DT is performing better than MLP for the $\tau$@5 and $\rho$@5. But in the case of full rank list, we observe MLP is outperforming DT and XGBoost for the $\tau$ and $\rho$.  \\ 
%\noindent\textbf{Duplicate question retrieval}: In the problem of duplicate question retrieval, we observe that our textual method is performing better than the other baseline textual procedures. Also, we have noticed that including node2vec based network features in our model brings significant improvement (~+1.63 in MRR) over the only textual features. The netMF based network features perform better (~+1.52 in MRR) than the only textual features. The improvement is statistically significant. In word2vec+MLP, We also observe that textual features+ network features are performing better than the only textual features, but the improvement is not statistically significant. In the case of the second baseline, i.e., doc2vec+MLP is not performing well at all for both the textual feature and text \& network features.

\section{Error analysis}
%\vspace{-0.2cm}
%~\rh{the discussion here is not providing any useful for the readers as the examples are missing and I assume only question ids are used (if that is the case, authors should have mentioned) [review]}
In this section, we test our models for various use cases to identify which variant of the model fails and when. Here, we demonstrate two use cases -- (a) \textbf{TE} performs better than \textbf{TE+network}: In Table~\ref{tab:erroranalysis} (up), we show a few test examples where \textbf{TE} performs better than \textbf{TE+network}. We observe that \textbf{TE} performs better when the title of the anchor question ($q_a$) is long and more detailed thus allowing the model to obtain a richer representation for the recommendation task. Even if the neighborhood of the most frequent tag of the anchor question is sparse, this does not affect the performance since the title text is elaborate and thus already rich in information. (b) \textbf{TE+network} performs better than \textbf{TE}: In Table~\ref{tab:erroranalysis} (down), we show few test examples where \textbf{TE+network} performs better than \textbf{TE}. \textbf{TE+network} performs better when the number of words in the title of the anchor question ($q_a$) is less but the size of the neighborhood of the most frequent tag of the anchor question is relatively high. This additional information from the network neighborhood compensates for the shorter length of the title text. This observation demonstrates how the network features could be effective in enhancing the overall performance of the model.

In Table~\ref{tab:comparison}, we present the predicted rankings for some of the most frequently asked questions.
We observe inclusion of network features improves the rank of the actual duplicate in the rank list compared to the state-of-the-art and our text-based model; however, existing models perform better in some contexts.
In the case of \textbf{DUPE} and \textbf{DupPredictor}, cosine similarity between titles, bodies, tags, and codes was generally used. Even then, cosine similarity scores as a feature do not help identify duplicates in a large ecosystem like Ubuntu since cosine similarity is more effective if the questions are either semantically similar or have a lot of word overlap.
\section{Conclusion and future work}
In this paper, we have proposed methods to solve the two CQA-related problems --(i) duplicate question retrieval and (ii) duplicate question confirmation time. In both problem statements, our model outperforms other state-of-the-art baseline models. Further adding network features, we obtained statistically significant improvements. In the future, we would like to investigate the temporal characteristics of questions that are closed as a duplicate. In addition, we would like to study other comparable datasets and tackle similar problems.

\bibliographystyle{plainnat}
\bibliography{references}

% \section*{References}

% References follow the acknowledgments. Use unnumbered first-level heading for
% the references. Any choice of citation style is acceptable as long as you are
% consistent. It is permissible to reduce the font size to \verb+small+ (9 point)
% when listing the references.
% {\bf Note that the Reference section does not count towards the eight pages of content that are allowed.}
% \medskip

% \small

% [1] Alexander, J.A.\ \& Mozer, M.C.\ (1995) Template-based algorithms for
% connectionist rule extraction. In G.\ Tesauro, D.S.\ Touretzky and T.K.\ Leen
% (eds.), {\it Advances in Neural Information Processing Systems 7},
% pp.\ 609--616. Cambridge, MA: MIT Press.

% [2] Bower, J.M.\ \& Beeman, D.\ (1995) {\it The Book of GENESIS: Exploring
%   Realistic Neural Models with the GEneral NEural SImulation System.}  New York:
% TELOS/Springer--Verlag.

% [3] Hasselmo, M.E., Schnell, E.\ \& Barkai, E.\ (1995) Dynamics of learning and
% recall at excitatory recurrent synapses and cholinergic modulation in rat
% hippocampal region CA3. {\it Journal of Neuroscience} {\bf 15}(7):5249-5262.

\end{document}